\documentstyle[sprocl]{article}

\input{psfig}

\def\be{\begin{equation}}
\def\ee{\end{equation}}
\def\bea{\begin{eqnarray}}
\def\eea{\end{eqnarray}}

\begin{document}

\title{UNIFIED PICTURE OF DIS AND DIFFRACTIVE DIS}

\author{C. Royon }

\address{CEA, DAPNIA, Service de Physique des Particules, \\
Centre d'Etudes de Saclay, France}


\maketitle\abstracts{The QCD dipole picture allows to build an unified 
theoretical description -based on BFKL dynamics- of the total and 
diffractive nucleon structure functions.
We use a four parameter fit to describe the 1994 H1
proton structure function $F_{2}$ data in the low $x$, moderate $Q^{2}$
range. Without any additional parameter, the gluon density and the
longitudinal structure functions are predicted. The diffractive dissociation
processes are discussed within the same framework,
and a new 6 parameter fit of the 1994 H1 data is performed which leads to a 
comprehensive description of $F_2^{D(3)}$.
}

\section{Introduction}
Considering the phenomenological discussion on the proton structure functions 
measured  by deep-inelastic scattering of
electrons and positrons at HERA, it is striking to realize that the proposed
models, on one side for the total quark structure function 
\cite{H1} $F_{2}\left(
x,Q^{2}\right) $ and on the other side for its diffractive component 
\cite{F2DH1} $%
F_{2}^{D\left( 3\right) }\left( x,M^{2},Q^{2}\right) $  are 
in general distinct.
\par
However, the quest for an unifying picture of total and
diffractive structure functions based on a QCD framework is a
challenge. The interest of using the QCD dipole approach \cite{Mueller}
for deep-inelastic structure functions is to deal with an unified approach based
on  the BFKL resummation
properties of perturbative QCD \cite{BFKL}.

\section{BFKL dynamics and the QCD dipole model}
To obtain the proton structure function $F_{2}$, we use the $k_{T}$ 
factorisation theorem \cite{catani}, valid 
for QCD at high energy (small $x$), 
in order to factorise the $(\gamma~ g(k) \rightarrow q ~ \bar{q})$ cross
section and the 
unintegrated gluon distribution of a proton containing the 
physics of the BFKL pomeron \cite{Mueller}. 
The detailed calculations can be found in \cite{ourpap}.
\par
We finally obtain:
\begin{eqnarray}
\label{predF2}
F_2 \equiv F_T + F_L 
= {\cal N} a^{1/2} e^{(\alpha_{P} -1) \ln\frac{c}{x}} \frac{Q}{Q_0}
e^{- \frac{a}{2} \ln^2 \frac{Q}{Q_0}} 
\end{eqnarray}
where
$\alpha_{P} -1 = \frac{4 \bar{\alpha} N_{C} \ln 2}{\pi}$, and
$a=\left(\frac{\bar{\alpha} N_c}{\pi} 7 \zeta(3) \ln\frac{c}{x}\right) ^{-1}$.
The free parameters for the fit of the H1 data are ${\cal N}$, $\alpha_{P}$, 
$Q_{0}$, and $c$.
Finally, we get $R$, and $F_{G}/F_{2}$,
which are independent of the overall normalisation ${\cal N}$ \cite{ourpap}
and represent parameter free predictions of the model once the $F_2$ fit
is performed.
\par
In order to test the accuracy of the $F_{2}$ parametrisation obtained in formula
(1), a fit using the recently published data from the H1 experiment
\cite{H1} has been
performed \cite{ourpap}.
We have only used the points with $Q^{2} \leq 150 GeV^{2}$ to remain
in the domain of validity of the 
QCD dipole model. The result of the fit is given in Figure 1.
The $\chi^{2}$ is 88.7
for 130 points, and the values of the parameters are 
$Q_{0}=0.522 GeV$, ${\cal N}= 0.059,$ and $c=1.750,$ while $\alpha_{P}=0.282$.
Commenting on the parameters, the effective coupling constant
 extracted from $\alpha_{P}$ is $\alpha =0.11$, close
to $\alpha (M_{Z})$ used in the H1 QCD fit. It is a small value for the 
fixed value of the coupling constant required by the BFKL framework.
The running
of the coupling constant is not taken into account in the present scheme. 
This could explain the rather low value of the effective $
\Delta _{P}$ which is expected to be decreased by the next leading 
corrections.
The value of $Q_{0}$ corresponds
to a tranverse size of 0.4 fm which 
is in the correct range for a proton non-perturbative characteristic
scale. 
\par
One obtains also a 
parameter-free prediction 
for $F_{L}$ \cite{papdiff} which is in agreement 
with the  (indirect)
experimental determination for  $%
F_{L}$ \cite{FLH1}, but somewhat lower than the center values. Thus, it would 
be interesting to 
obtain a more precise measurement of $F_{L}$ to
test the different predictions on the $Q^2$-evolution 
as already mentionned in Ref. \cite{ourpap}.

\begin{figure}
\begin{center}
\psfig{figure=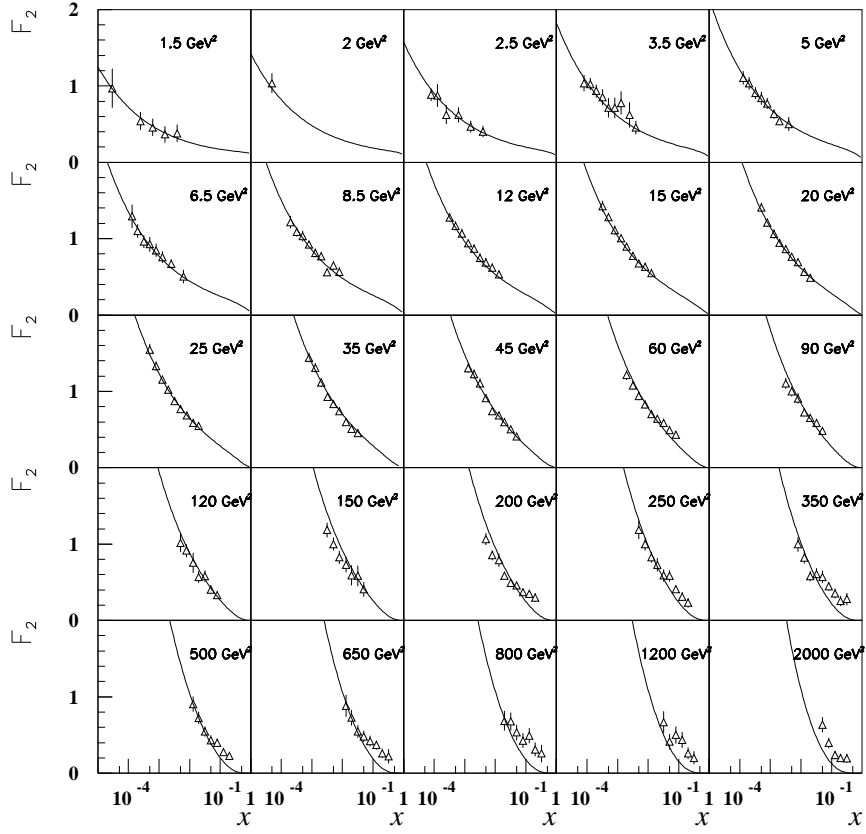,height=5in}
\end{center}
\caption{Results of the 4-parameter fit of the H1 proton
structure function (the fit has been performed using the points with
$Q^{2} \leq 150 GeV^{2}$). We note a
discrepancy at high $x$, high $Q^{2}$ due, in particular, to the absence of
the valence contribution
not considered  in the present model. 
\label{fig:radi}}
\end{figure}

\section{Diffractive structure functions}

The success of the dipole model applied to the proton structure function
motivates its extension to the investigations to other inclusive processes,
in particular
to diffractive dissociation. We can distinguish two different components:
\newline
- the ``elastic" term which represents the elastic scattering of the onium
on the target proton;
\newline
- the ``inelastic" term 
which represents the sum of all dipole-dipole
interactions (it is dominant at large masses of the excited system).
\newline
Let us describe in more details each of the two components. The
``inelastic"
term dominates at low $\beta$, where $\beta=x/x_{P}$, 
$x_{P}$ is the proton momentum fraction carried by the
colourless exchanged object \cite{inelcomp}.
This component, integrated over $t$, the momentum transfer, is factorisable
in a part depending only on $x_p$ (flux factor) and a factor depending only on
$\beta$ and $Q^{2}$ (``pomeron" structure function) \cite{inelcomp}.
\begin{eqnarray}
F_{2}^{D(3)} (Q^{2},x_p,\beta)  = \Phi (x_{p}) F_{P}(Q^{2},\beta) 
\end{eqnarray}
where $a$ is defined in the first chapter, which gives the following
formula for the inelastic component after a saddle point approximation:
\begin{eqnarray}
F_{T,L}^{D(3),inel.} \simeq
\frac{1}{x_P}\frac{Q}{Q_0}a^3(x_P)\;x_P^{-2\Delta_P}
\sqrt{a(\beta)}\beta^{-\Delta_P} e^{-\frac{a(\beta)}{2}\log^2\frac{Q}{4Q_0}}
\end{eqnarray}
 
The important point to notice is that
$a(x_{P})$ is proportional of $ln (1/x_{P})$. The effective exponent
(the slope of $ln F_{2}^{D}$ in $ln x_{P}$) is found to be dependant
on $x_{P}$ because of the term in $ln^{3}(x_{P})$ coming from $a$,
and is sizeably smaller than the BFKL exponent. This is why we can
describe an apparently softer behaviour with
the BFKL equation, which predicts a harder behaviour (the formal exponent in
$x_{P}$ is close to 0.35). This is due to the fact that the effective exponent
is smaller than the formal one. 
\par
For the elastic component, one considers
\begin{eqnarray}
F_T^{D(3),el.} \simeq \frac{1}{x_P}
a^3(x_P)\log^3\frac{Q}{2Q_0\sqrt{\beta}}\;
x_P^{-2\Delta_P}\;e^{-a(x_P)\log^2\frac{Q}{2Q_0\sqrt{\beta}}}\\
\times\beta(1\!-\!\beta)\;
\left[{~_2F_1}\left(-\frac{1}{2},\frac{3}{2};2;1-\beta\right)
\right]^2
\end{eqnarray}
and adds ${F_{L}^{D(3),qel.}}$ where $ F_{L}$ is obtained from the previous
formula by changing
${~_2F_1}\left(-\frac{1}{2},\frac{3}{2};2;1-\beta\right)$
to ${~_2F_1}\left(-\frac{1}{2},\frac{3}{2};1;1-\beta\right)$.
The elastic component behaves quite differently \cite{inelcomp} from the
inelastic one.
First it dominates at
$\beta \sim 1$. It is also factorisable like the inelastic component, but
with a different flux factor, which means that the sum of the two
components will not be factorisable. This means that in this model,
factorisation breaking is coming partially from the fact that we sum up two 
factorisable components with different flux factors.
The $\beta$ dependence is quite
small at large $\beta$, due to the interplay
between the longitudinal and transverse components. The sum remains almost
independent of $\beta$, whereas the ratio $R=F_{L}/F_{T}$ is strongly
$\beta$ dependent. Once more, a $R$ measurement in
diffractive processes will be an
interesting way to distinguish the different models, as the dipole model
predicts specific $\beta$ and $Q^{2}$ behaviours for $F_L$ and $F_T$.
\par
We can then compare the H1 data to the following prediction:
\begin{equation}
F_2^{D(3)} = F_2^{D(3),inel}+F_2^{D(3),qel}+F_2^{D(3),Reggeon}
\end{equation}
where  
$F_2^{D(3),Reggeon}$ represents a secondary trajectory and 
is taken as in Reference \cite{F2DH1}.
The free parameters used in the fit are $\alpha_P(0)$, $\alpha_R(0)$, 
respectively the
exponents for the QCD pomeron and the secondary reggeon, the normalisations in
front of the elastic, inelastic and the reggeon terms, and the $Q_0$
parameter. The $\chi^2$ value obtained is quite good (305.5 for 220 degrees
of freedom with statistical error only, 243 for statistical and systematical
errors added in quadrature). The values of the parameters are the following:
$\alpha_P(0)=1.37$, $\alpha_R(0)=0.53$ (which is a coventional value for the
secondaries), $Q_0=0.43$ GeV. The values of $\alpha_P(0)$, and $Q_0$ are
different from those obtained with the $F_2$ fit because we used so far
approximate formulas to describe diffraction, and because the NLO corrections
can behave differently for total and diffractive DIS.
The results are given in Figure 2.

\begin{figure}
\vskip 3cm
\psfig{figure=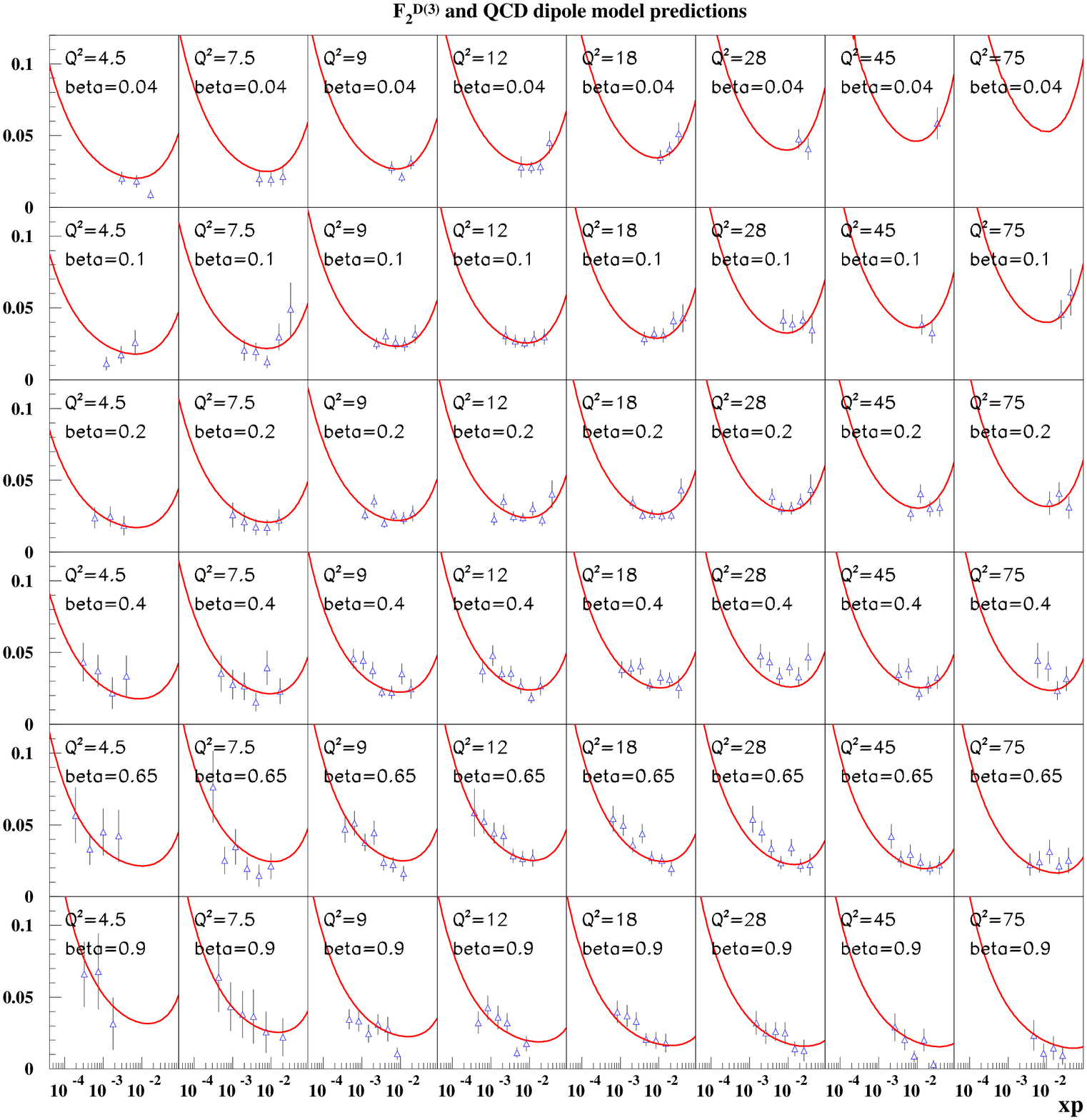,height=5in}
\caption{Result of the $F_2^{D(3)} fit$
(cf text) $~~~~~~~~~~~~~~~~~~~~~~~~~~~~~~~~~~~$
\label{fig:rad3}}
\end{figure}

\section*{Acknowledgments}
The results described in the present contribution come from a fruitful
collaboration with A.Bialas, S.Munier, H.Navelet, and R.Peschanski.
I also thank R.Peschanski 
for reading and comments on the manuscript.

\end{document}